\documentclass[conference]{IEEEtran}
\IEEEoverridecommandlockouts
\pdfoutput=1
\usepackage{cite}
\usepackage{amsmath,amssymb,amsfonts}
\usepackage{algorithmic}
\usepackage{graphicx}
\usepackage{textcomp}
\usepackage{xcolor}

\usepackage{multicap}
\usepackage{subfig}
\usepackage{graphicx}
\usepackage{epstopdf}
\usepackage{hyperref}
\usepackage[ruled,vlined]{algorithm2e}
\usepackage{multicol}
\usepackage{multirow}
\usepackage{array}

\def\BibTeX{{\rm B\kern-.05em{\sc i\kern-.025em b}\kern-.08em
    T\kern-.1667em\lower.7ex\hbox{E}\kern-.125emX}}
\begin{document}

\title{Boosting the Performance of Degraded Reads \\in RS-coded Distributed Storage Systems
}

\author{\IEEEauthorblockN{Tian Xie\IEEEauthorrefmark{2},
		Juntao Fang\IEEEauthorrefmark{1},
		Shenggang Wan\IEEEauthorrefmark{2} and 
		Changsheng Xie\IEEEauthorrefmark{1}}
	\IEEEauthorblockA{\IEEEauthorrefmark{1}Wuhan National Laboratory for Optoelectronics, Huazhong University of Science and Technology, China}
	\IEEEauthorblockA{\IEEEauthorrefmark{2}School of Computer Science and Technology, Huazhong University of Science and Technology, China}
	\ tian.xie.chn@gamil.com,\{yydfjt, sgwan, cs\_xie\}@hust.edu.cn}

\maketitle

\begin{abstract}
Reed-Solomon (RS) codes have been increasingly adopted by distributed storage systems in place of replication, because they provide the same level of availability with much lower storage overhead. However, a key drawback of those RS-coded distributed storage systems is the poor latency of degraded reads, which can be incurred by data failures or hot spots, and are not rare in production environments. To address this issue, we propose a novel parallel reconstruction solution called APLS. APLS leverages all surviving source nodes to send the data needed by degraded reads and chooses light-loaded starter nodes to receive the reconstructed data of those degraded reads. Hence, the latency of the degraded reads can be improved. Prototyping-based experiments are conducted to compare APLS with ECPipe, the state-of-the-art solution of improving the latency of degraded reads. The experimental results demonstrate that APLS effectively reduces the latency, particularly under heavy or medium workloads.
\end{abstract}

\begin{IEEEkeywords}
Degraded Read; Parallelism; Erasure Code; Storage System
\end{IEEEkeywords}

\section{Introduction}
\label{sec:it}
Reed-Solomon (RS) coded distributed storage systems are seeing an increasing popularity in modern data centers, because they provide high availability at low storage overhead~\cite{OSDI10FORD-AVAILABILITY, HOTSTORAGE13Rashmi-warehouse, atc17-giza}. In those distributed storage systems, which consist of thousands of nodes and service millions of requests per second, data failures and hot spots are not rare~\cite{OSDI10FORD-AVAILABILITY}. Poor latency degraded reads are induced when these failed data and hot data are requested.

To apply an arbitrary RS(k,m) code in distributed storage systems, files are first divided into fixed size data chunks. Then, $m$ parity chunks are generated from these $k$ data chunks through the arithmetic over Galois Fields~\cite{Plank1997A}. All those $k+m$ data and parity chunks together form a stripe which can tolerate up to $m$ internal lost chunks by itself. At last, the $k+m$ data and parity chunks in that stripe are distributed to $k+m$ storage nodes.

In tradition, when a chunk becomes unavailable (due to the data failures or hot spots), it is reconstructed by $k$ out of the $k+m-1$ surviving chunks. First, a storage node in the distributed storage system is assigned as a starter node. Then, $k$ arbitrary surviving chunks in the same stripe of the unavailable chunk are fetched from $k$ source nodes (storage nodes containing surviving chunks needed by the reconstruction) to that starter node. Finally, the starter node uses the $k$ surviving chunks to reconstruct the unavailable chunk using the Galois Field arithmetic. All the above procedures are known as a \emph{degraded read}. Compared to reading an available chunk directly, which is known as a \emph{normal read}, the degraded read suffers from a much longer latency due to that increasing data transmission ($1$ chunk vs. $k$ chunks).

To improve the latency of the degraded read, solutions are proposed to explore its internal parallelism. In tradition, the $k$ surviving chunks are sent by $k$ source nodes, but only one starter node receives them. Clearly, the downstream network bandwidth of the starter node is the bottleneck of the degraded read. Consequentially, via deploying some agent nodes between the source nodes and the starter node to receive the surviving chunks and do the reconstruction in parallel, PPR~\cite{Eurosys2016PPR} and ECPipe~\cite{ATC17-lirepair} reduce the amount of data that should be received by the starter node and thus improve the latency of the degraded read.

However, even being optimized by the aforementioned solutions, the latency gap between degraded reads and normal reads is still significant. For example, to read a 4MB data in ECPipe~\cite{ATC17-lirepair}, the latency of degraded reads is 1.3-1.6x to that of normal reads in the best cases. In addition, this latency gap is hard to be improved through deploying more agent nodes, because it is usually determined by the inherent features of networks, e.g., hop latencies and imbalanced parallelism.

Nevertheless, by our analysis, that latency gap can still be closed, which roots in the following three dedicated observations on RS-coded distributed storage systems.

\begin{figure*}[t]
	\small
	\centering	
	\subfloat[Traditional method: Four source nodes paticipate in the degrade read. $S_4$ is chosen as the starter node. $3\times c$ data is sent to $S_4$, the requested data is reconstructed on $S_4$.]
	{
		\label{fig:11}
		\includegraphics[width=.44\textwidth]{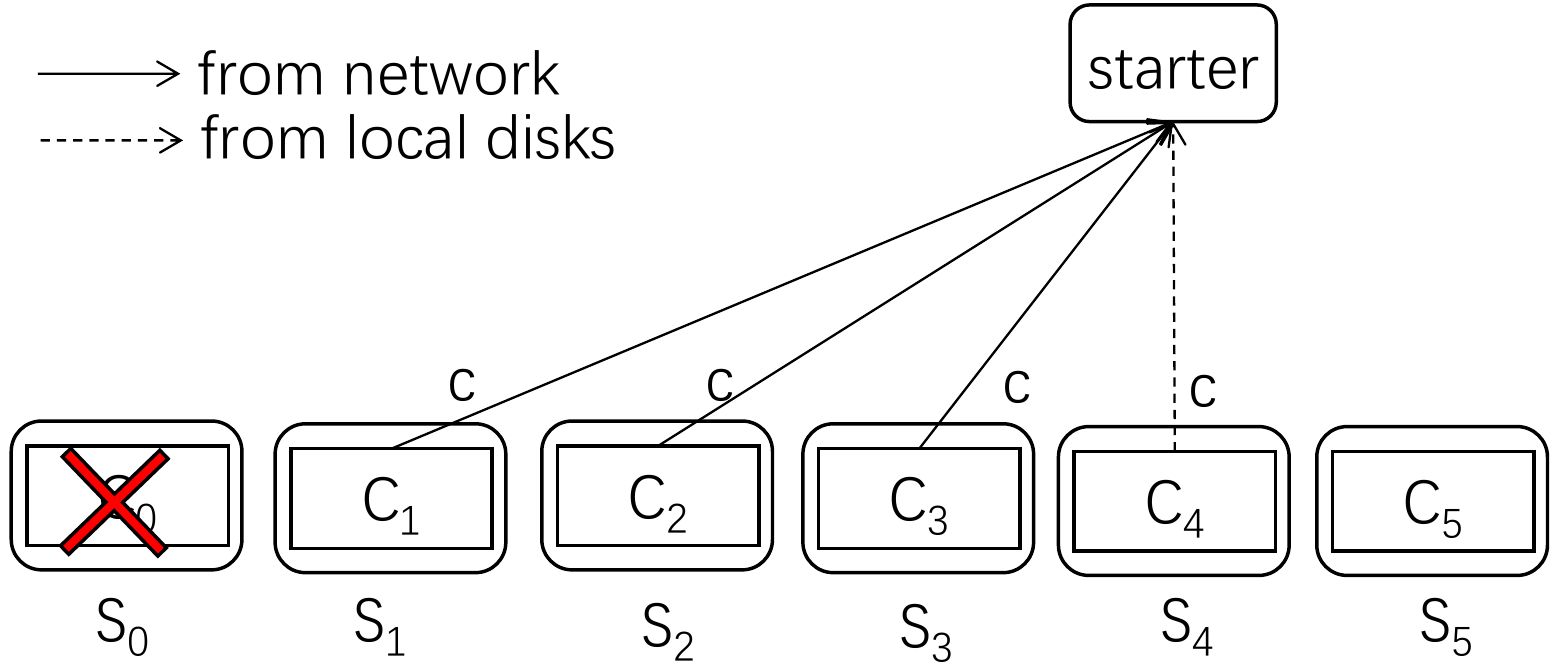}
		
	}
	\hspace{4mm}
	\subfloat[APLS method: Five source nodes paticipate in the degrade read. a light-loaded storage node is chosen as the starter node, $S_i$ ($i = 1, 2, .... ,5$) are agent nodes. Each agent node receives $\frac{3\times c}{5}$ from other three source nodes, reconstructs $\frac{c}{5}$ data and sends the reconstructed data to the starter node.]
	{
		\label{fig:12}
		\includegraphics[width=.44\textwidth]{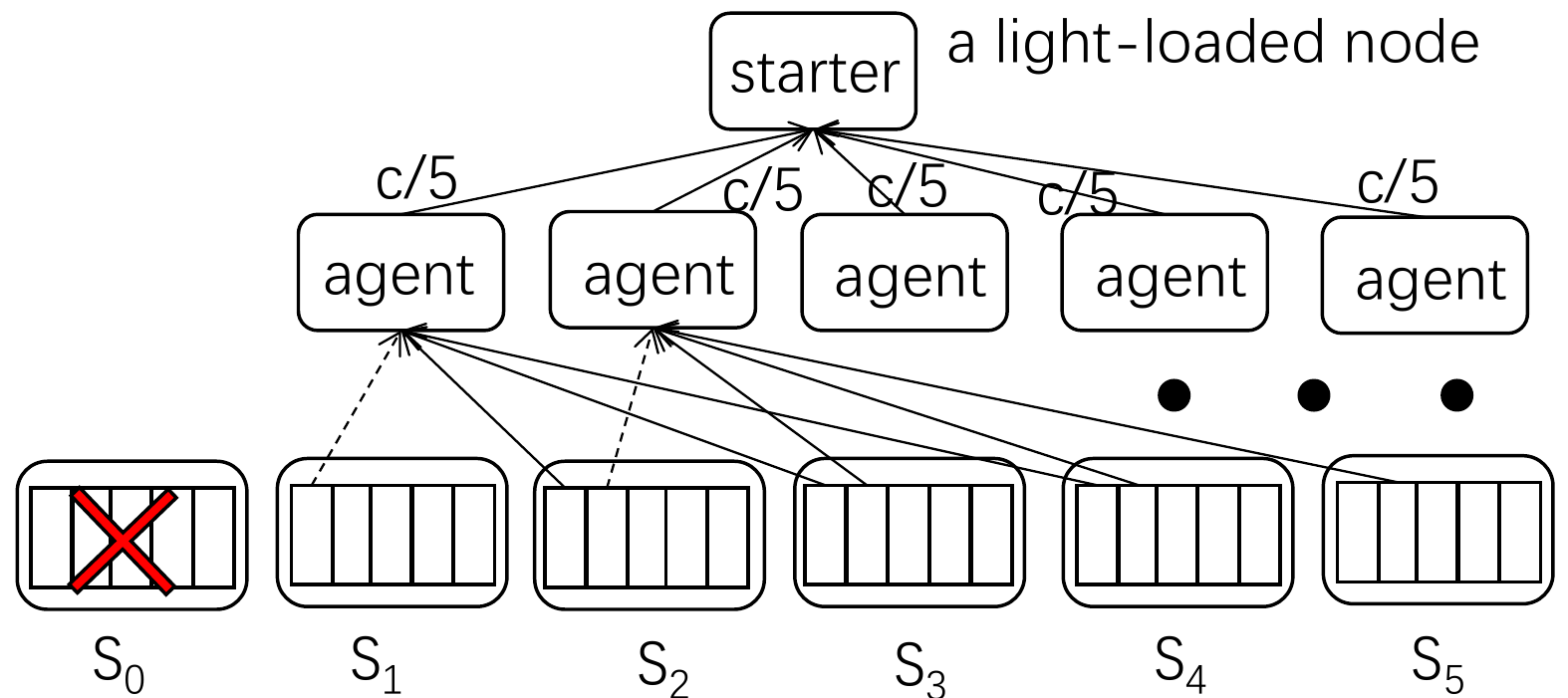}
	}
	\caption{Examples of the degraded read using the traditional method and APLS in an RS(4,2)-coded distributed storage system. A stripe has six chunks: $C_i$ ($i = 0, 1,.... ,5$). $C_i$ is stored on $S_{i}$ ($i = 0, 1, .... ,5$). The chunk size is $c$. }
	\label{fig:dro}
\end{figure*}
\normalsize

\textbf{Obs. 1: Optimized by the state-of-the-art parallel reconstruction approaches, e.g., ECPipe~\cite{ATC17-lirepair}, a dedicated balance has been reached between the data sending on each source node and the data receiving on the starter node}. When those approaches are applied, the starter node only needs to receive one chunk. Because $k$ chunks are sent by $k$ source nodes, the times for sending and receiving data are roughly the same under a uniform reconstruction bandwidth assumption that all nodes have the same downstream and upstream network bandwidth for reconstruction.

\textbf{Obs. 2: The downstream network bandwidth for receiving data on the starter node can be further improved through choosing a light-loaded starter node}. As we mentioned in Obs. 1, existing approaches follow the uniform reconstruction bandwidth assumption. However, in large scale distributed storage systems, some nodes actually have much more available bandwidth than other nodes because of the imbalanced workload distribution~\cite{Atikoglu12,googledata13,Huang14,Novakovic16,Lu2017Imbalance}. If a light-loaded storage node is assigned as the starter node of a degraded read, more network bandwidth is available to receive the same amount of data (one chunk).

\textbf{Obs. 3: More source nodes can participate in the reconstruction of a chunk by performing the reconstruction in a smaller unit than the chunk, thus improving the total network bandwidth of source nodes}. Although data are organized in chunks, words which are a few to tens of bits are actually the basic unit for RS coding~\cite{Plank1997A}, i.e., a chunk usually consists of millions of words. In addition, an unavailable word can be reconstructed by arbitrary $k$ out of the rest $k+m-1$ available words. Hence, if words in an unavailable chunk are reconstructed by words from different $k$-chunk combinations, up to $k+m-1$ source nodes can provide the surviving data concurrently, while keeping the amount of the surviving data to be sent unchanged. In other words, more network bandwidth can be achieved for sending data.

Through increasing the network bandwidth for both sending the surviving data and receiving the reconstructed data, which is probably the bottleneck of degraded reads, the latency of degraded reads can be further improved, thus closing the gap between normal reads and degraded reads.

The contributions of our work are summarized as follows.

(1) We propose APLS (\emph{All Parallelism with a Light-loaded Starter, reads as A Plus}) to improve the latency of degraded reads in RS-coded distributed storage systems through exploring and exploiting the potential network bandwidth for transmitting data for reconstruction. More specifically, APLS uses all surviving source nodes to increase the network bandwidth for sending the data of reconstruction, and chooses light-loaded starter nodes to increase the network bandwidth for receiving the reconstructed data. Hence, the degraded reads can be speeded up.

(2) We build a model and conduct comprehensive theoretical analysis to quantitatively evaluate APLS. The analysis results demonstrate the effectiveness and efficiency of APLS.

(3) Besides the theoretical analysis, a prototype of APLS is built upon a storage cluster consisting of sixteen servers. Experimental results validate our theoretical analysis. For example, compared to the state-of-the-art solution, ECPipe, the latency of degraded reads is reduced by up to 28\% via using APLS when nodes have medium or heavy workloads.

The rest of this paper is organized as follows. Section~\ref{sec:bgmv} introduces the background materials and motivations. Section~\ref{sec:de} details the design of APLS. Section~\ref{sec:pe} presents the prototype and its results to demonstrate the effectiveness of APLS on the latency of degraded reads. Section~\ref{sec:rw} reviews some related work in the literature and Section~\ref{sec:con} concludes this paper.

\section{Background and Motivation}
\label{sec:bgmv}
In this section, we first review background materials of Reed-Solomon (RS) coding and optimizations on degraded reads, and then illustrate our motivations.

Terms to facilitate our discussion are summarized as follows.

\textbf{\emph{Word}}: A word is the basic unit to build an RS code. It typically ranges from a single bit to a few bytes~\cite{Plank1997A}.

\textbf{\emph{Chunk}}: A chunk is the basic unit for data placement in distributed storage systems. Typically, a chunk is a set of continuous words. In addition, it can be a \emph{data chunk} or \emph{parity chunk}.

\textbf{\emph{Stripe}}: A stripe is an independent group of data and parity chunks. An RS(k,m)-coded stripe has $k$ data chunks and $m$ parity chunks. In some cases, lost chunks can be reconstructed by surviving chunks in their hosting stripe.

\textbf{\emph{Packet}}: A packet is the basic unit for network transmission. The packet size is typically larger than the word size, and can be configured.

\textbf{\emph{Normal read}}: An operation to read available data directly.

\textbf{\emph{Degraded read}}: An operation to read unavailable but recoverable data through RS decoding.

\textbf{\emph{Source node}}: In a degraded read, a source node is a node which contains data for reconstructing the requested data. In RS(k,m)-coded distributed storage systems, there are at most $k+m-1$ source nodes during the degraded read. In existing solutions, only $k$ source nodes are used; in contrast, in APLS, up to $k+m-1$ source nodes are used.

\textbf{\emph{Starter node}}: In a degraded read, a starter node is a node which initiates the degraded read. In existing solutions, the starter node is chosen from source nodes; in APLS, we choose a light-loaded storage node as the starter node.

\begin{figure*}[t]
	\small
	\centering	
	\subfloat[An RS(4,2)-coded stripe consists of four data chunks and two parity chunks.]
	{
		\centering
		\label{fig:30}
		\includegraphics[width=.25\textwidth ]{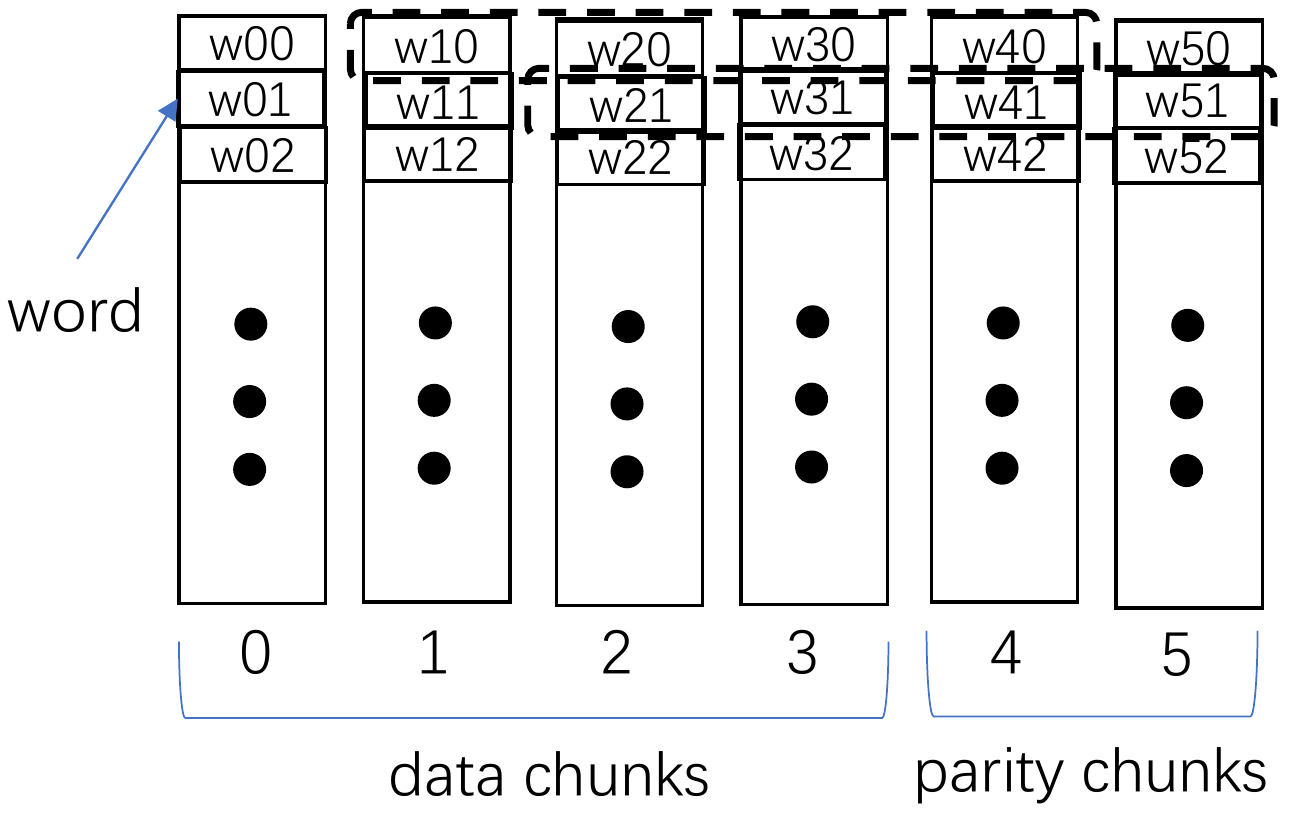}
		
	}
	\subfloat[RS(4,2) encoding.]
	{
		\label{fig:encode}
		\includegraphics[width=.30\textwidth]{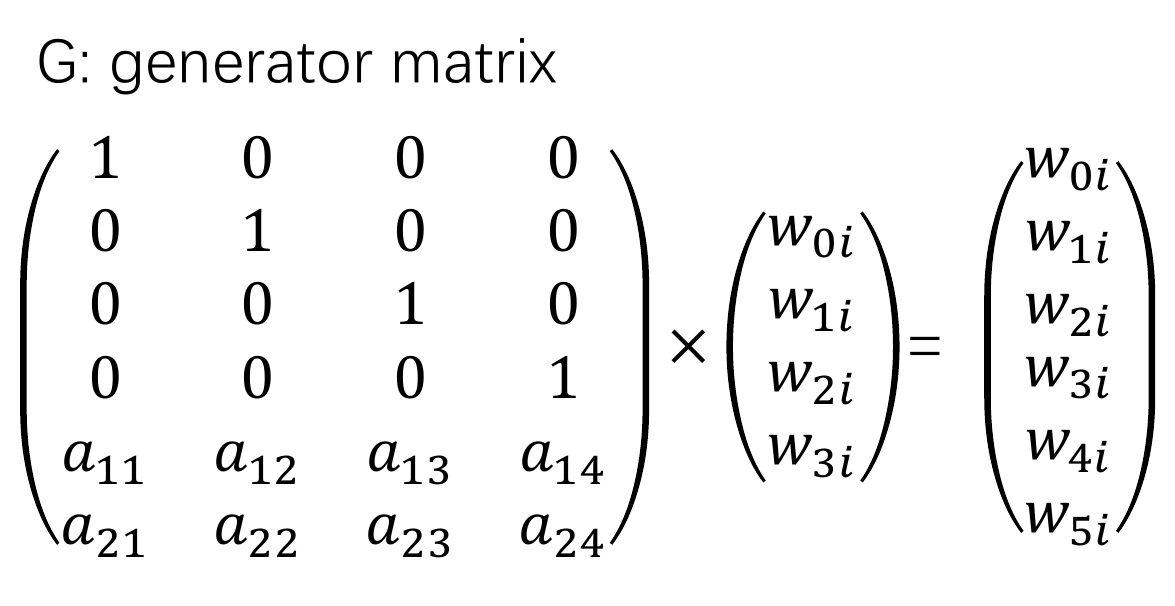}
		
	}
	\subfloat[RS(4,2) decoding.]
	{
		\label{fig:decode}
		\includegraphics[width=.40\textwidth]{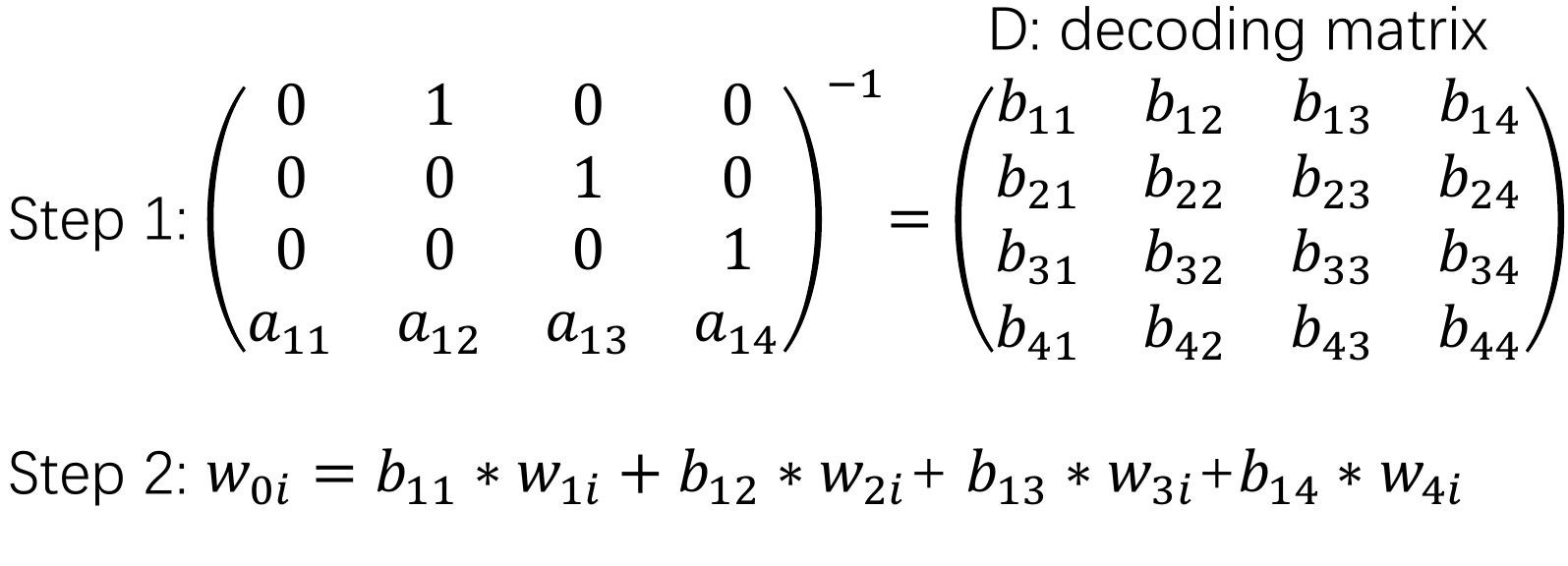}
	}
	\caption{Encoding and decoding in an RS(4,2) code.}
	\label{fig:ende}
\end{figure*}
\normalsize

\begin{figure*}[t]
	\small
	\centering	
	\subfloat[PPR method: $S_4$ is chosen as the starter node, $S_2$ is chosen as an agent node. First, $b_{11}\times w_{1i}$'s are sent from $S_1$ to $S_2$, and $b_{13}\times w_{3i}$'s are sent from $S_3$ to $S_4$; second, $\sum_{j=1}^{2} b_{1j}\times w_{ji}$'s are sent from $S_2$ to $S_4$, the requested data are reconstructed on $S_4$.]
	{
		\label{fig:p2}
		\includegraphics[width=.44\textwidth]{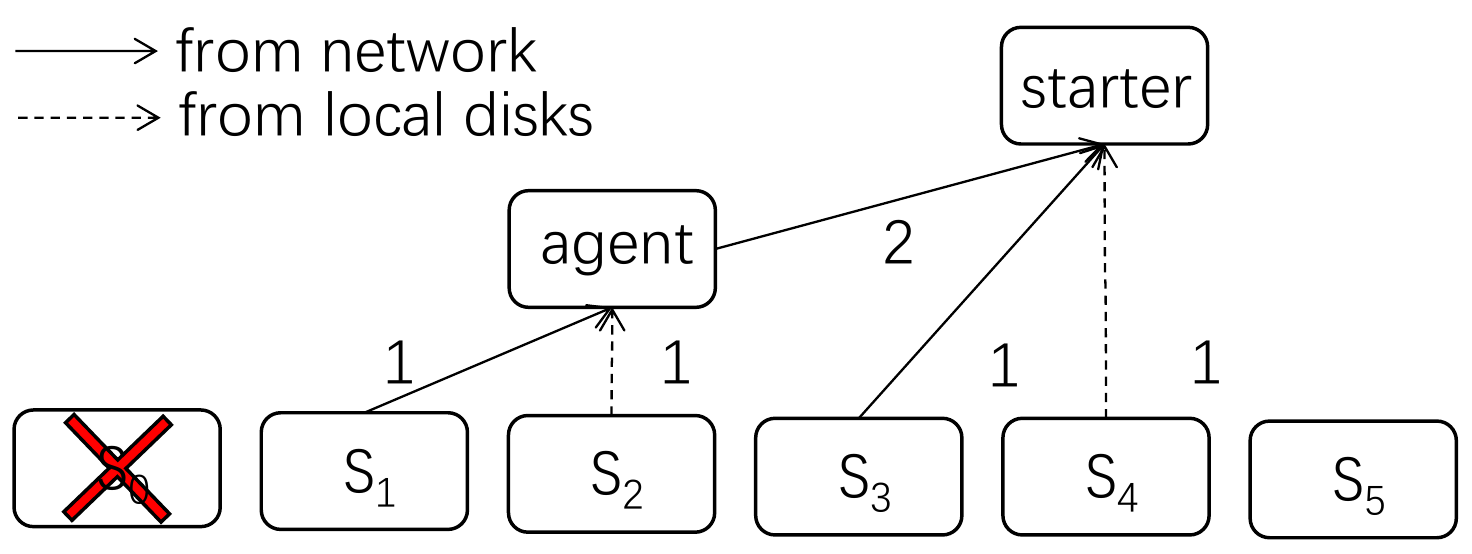}
	}
	\hspace{4mm}
	\subfloat[ECPipe method: $S_4$ is chosen as the starter node, $S_2$ and $S_3$ are chosen as agent nodes. Each word is recontruted by three steps. First, $b_{11}\times w_{1i}$ is sent from $S_1$ to $S_2$; second, $\sum_{j=1}^{2} b_{1j}\times w_{ji}$ is sent from $S_2$ to $S_3$; third, $\sum_{j=1}^{3} b_{1j}\times w_{ji}$ is sent from $S_3$ to $S_4$, the requested data are reconstructed on $S_4$. Different words are recontrutectd using pipelining.]
	{
		\label{fig:p3}
		\includegraphics[width=.44\textwidth]{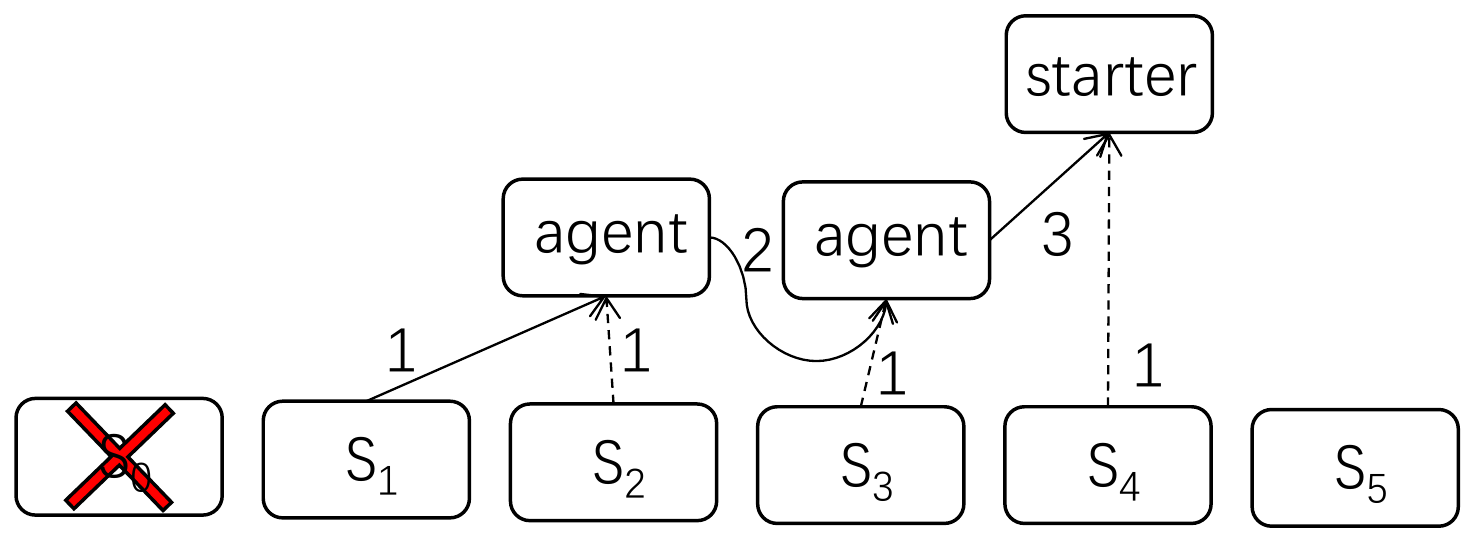}
	}
	\caption{Examples of the degraded read using PPR and ECPipe in an RS(4,2)-coded distributed storage system. The conditions are similar to Figure~\ref{fig:dro}. $w_{ji}$'s are words of chunk $j$, and $b_{1j}$'s are decoding coefficients.}
	\label{fig:dr}
\end{figure*}
\normalsize

\textbf{\emph{Agent node}}: In a degraded read, an agent node is a node which receives the surviving data, performs decoding computation, and sends data to other agent nodes or the starter node. The agent nodes are chosen from source nodes to reduce network transmission in both the existing solutions and APLS.

\textbf{\emph{Upstream Network Bandwidth}}: Upstream network bandwidth refers to the speed in which data are transferred out over the network. In a degraded read, the upstream network bandwidth limits the data sent by source nodes.

\textbf{\emph{Downstream Network Bandwidth}}: Downstream network bandwidth refers to the speed in which data are transferred in over the network. In a degraded read, the downstream network bandwidth limits the data received by the starter node.

Symbols used in this paper are listed in Table~\ref{tab:sd}.

\begin{table}[t]
	\footnotesize
	\centering
	\caption{Symbols and Their Definitions}
	\begin{tabular}{|c|c|}
		\hline
		\textbf{Symbol} & \textbf{Definition}\\
		\hline
		\multirow{1}{*}{$k$} 	& \# of data chunks in a stripe\\
		\hline
		\multirow{1}{*}{$m$} 	& \# of parity chunks in a stripe\\
		\hline
		\multirow{1}{*}{$c$} 	&  Size of a chunk (bytes)\\
		\hline
		\multirow{2}{*}{$q$} 	& \# of source nodes which are used \\ &during a degraded read operation\\
		\hline
		\multirow{1}{*}{$B$}	& Network bandwidth on each node\\
		\hline
		\multirow{2}{*}{$\theta_s$}	& Ratio of network bandwidth used for the degraded read\\
		&to the total network bandwidth on source nodes  \\
		\hline
		\multirow{2}{*}{$T_{dr}(c)$} & The latency of the degraded read to \\
		& data with size of c \\
		\hline
	\end{tabular}
	\label{tab:sd}
\end{table}
\normalsize

\subsection{Reed-Solomon Coding}
\label{subsec:rsc}
RS codes~\cite{Reed1960Polynomial} are a set of popular erasure codes which are widely used in real world distributed storage systems~\cite{SOSP03GHEMAWAT-GFS, OSDI10FORD-AVAILABILITY, USENIX12HUANG-LRC, HOTSTORAGE13Rashmi-warehouse, vldb2013qfs,FAST14Tyler-hdfs-facebook} to ensure data availability. In those distributed storage systems, such as GFS~\cite{SOSP03GHEMAWAT-GFS}, HDFS~\cite{FAST14Tyler-hdfs-facebook} and WAS~\cite{USENIX12HUANG-LRC}, small files are accumulated into large files, and large files are divided into the data chunks with a fixed chunk size ranging from 4 MB~\cite{atc17-giza} to 256 MB~\cite{HOTSTORAGE13Rashmi-warehouse}.

Although the data and parity are organized as chunks in RS-coded distributed storage systems, words which are composed of a few bits are indeed the basic unit of RS coding. To apply RS coding, each data chunk is first divided into words. And then, words of parity chunks are calculated from words at the same offset of each data chunk of the same stripe based on Galois Field arithmetic~\cite{Plank1997A}. For example, in Figure~\ref{fig:encode}, each word $w_{4i}$ in chunk 4 is constructed by a matrix multiplication where words $w_{0i}$, $w_{1i}$, $w_{2i}$, and $w_{3i}$ are multiplied by a generator matrix G. Assume a word is composed of $w$ bits, the elements of the matrix G are numbers in $GF(2^w)$.

When a data chunk is unavailable in RS(k,m)-coded storage systems, its reconstruction consists of two steps. The first step is calculating a decoding matrix D by taking the inverse of a matrix created using any $k$ surviving rows of G. And the second step is reconstructing the chunk by reconstructing all its words. Each word $w_i$ of the chunk is reconstructed via Equation~\ref{eqa:decode} where $d_j$'s are corresponding elements of the decoding matrix D and $w_{ji}$'s are corresponding words.

\vspace{0pt}
\begin{equation}
\label{eqa:decode}
w_i = \sum_{j=1}^{k}d_{j}\times w_{ji}
\end{equation}

For example, as shown in Figure~\ref{fig:decode}, chunk 0 is unavailable. By choosing the second, third, fourth and fifth rows of G, we get a $4\times 4$ matrix. By reversing this matrix, we get the decoding matrix D. The decoding coefficients are elements of the first row of D. Finally, any word $w_{0i}$ in chunk 1 is reconstructed by: $w_{0i} = b_{11}\times w_{1i} + b_{12}\times w_{2i} + b_{13}\times w_{3i} + b_{14}\times w_{4i}$.

\subsection{Optimizations on Degraded Reads}
\label{subsec:odr}
In tradition~\cite{USENIX12HUANG-LRC, Hu2017Latency}, to reduce the network transmission, a source node which contains the surviving chunk in the same stripe of the requested chunk is chosen as the starter node of a degraded read. The starter node fetches $k-1$ chunks from other source nodes, and then reconstructs that requested chunk. In such a case, the starter node is highly congested by receiving the data. As shown in Figure~\ref{fig:11}, $S_4$ is the starter node and three chunks are sent to $S_4$ to reconstruct the unavailable chunk. Therefore, the downstream network bandwidth of $S_4$ is the bottleneck of the degraded read.

Two types of methods are proposed to reduce the latency of degraded reads by exploring the parallelism inside those degraded reads. One is exploiting the parallelism inside reconstructing one single word, i.e., doing the multiplication operations in Equation~\ref{eqa:decode} on different nodes. For example, PPR~\cite{Eurosys2016PPR} decomposes the reconstruction operation into multiple partial operations (e.g., $b_{11}\times w_{1i}$ is calculated on $S_1$ in Figure~\ref{fig:p2}.) and leverages multiple agent nodes to perform the partial operations in parallel (e.g., $b_{11}\times w_{1i}$ and $b_{13}\times w_{3i}$ are calculated in parallel on $S_1$ and $S_3$, respectively, in Figure~\ref{fig:p2}.). As shown in Figure~\ref{fig:p2}, $S_4$ is the starter node, and $S_2$ is the agent node, two chunks are sent to $S_4$, i.e., the downstream network bandwidth of $S_4$ is still the bottleneck.

The other is leveraging the parallelism among the reconstruction of multiple words, i.e., reconstructing different words in an unavailable chunk on different nodes. For example, ECPipe~\cite{ATC17-lirepair} proposes to distribute the partial operations for reconstruction on more agent nodes and further do the reconstruction in parallel (e.g., three steps are done using pipelining in Figure~\ref{fig:p3}.). As shown in Figure~\ref{fig:p3}, $S_4$ is the starter node, and $S_2$ and $S_3$ are agent nodes, one chunk is sent to $S_4$. As a result, degraded reads can achieve the same performance as normal reads in theory.

According to the above analysis, existing optimizations focus on exploring the parallelism inside the computation procedure of degraded reads, but highly neglecting the parallelism among all surviving source nodes.

\subsection{Motivation}
\label{subsec:moti}
In distributed storage systems, failures could occur frequently for various reasons, including hardware/software failures, network connection loss, software bugs, crashes, updates, and system reboots~\cite{OSDI10FORD-AVAILABILITY, HOTSTORAGE13Rashmi-warehouse}. The failures often lead to degraded reads. Degraded reads are a critical concern for RS-coded distributed storage systems because they suffer from a much longer latency compared to normal reads. The poor latency of degraded reads can significantly affect user experience, and even reduce the revenue of service providers~\cite{2016outage}.

Therefore, various methods~\cite{Eurosys2016PPR, ATC17-lirepair} have been proposed to improve the performance of degraded reads in RS-coded storage systems by exploiting the reconstruction parallelism inside the degraded read. Based on the analysis made in Section~\ref{subsec:odr}, the latency of the degraded read is limited by both the upstream network bandwidth of source nodes (except the starter node) and the downstream network bandwidth of the starter node. Therefore, the latency of the degraded read cannot be further reduced even if more agents nodes are used.

However, the latency of degraded reads can be further reduced due to the following two reasons.

First, existing solutions only use $k$ source nodes during the degraded read. For example, the source node $S_5$ is not used by existing methods as shown in Figure~\ref{fig:dr}. The upstream network bandwidth of source nodes can be increased if more source nodes are used.

\begin{figure}[t]
	\small	
	\centering
	\includegraphics[width=.42\textwidth]{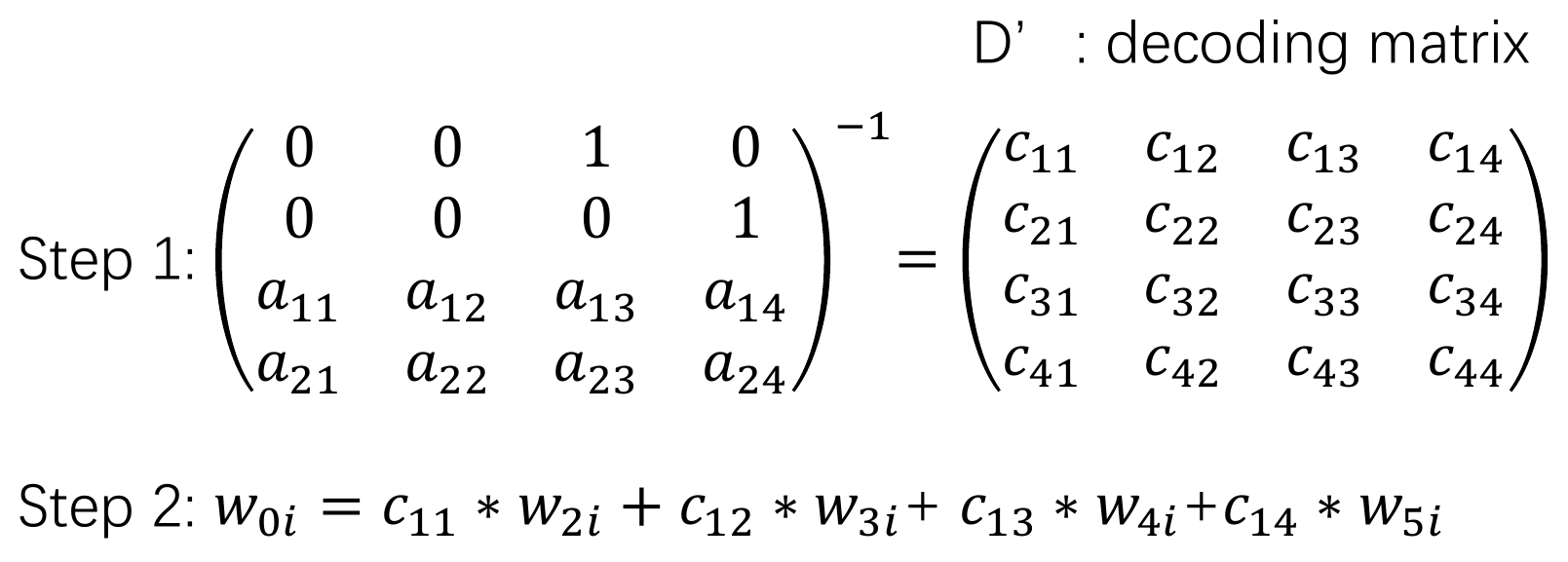}
	\caption{RS(4,2) decoding.}
	\label{fig:decode1}
\end{figure}
\normalsize

In RS(k,m)-coded distributed storage systems, an unavailable word can be reconstructed by words from arbitrary $k$ surviving chunks in the same stripe. For example, in Figure~\ref{fig:decode1}, chunk 0 is unavailable, by choosing the third, fourth, fifth and sixth rows of G, we get another decoding matrix D'. Then, a word $w_{0i}$ in chunk 0 is reconstructed by: $w_{0i} = c_{11}\times w_{2i} + c_{12}\times w_{3i} + c_{13}\times w_{4i} + c_{14}\times w_{5i}$.

Therefore, words of one unavailable chunk can be reconstructed by words from different $k$-chunk combinations. For example, as shown in Figure~\ref{fig:30}, to reconstruct words $w_{11}$ and $w_{12}$ in chunk 1, the word $w_{11}$ is reconstructed by words from chunks 2,3,4, and 5; the word $w_{12}$ is reconstructed by words from chunks 3,4,5, and 6. Therefore, all surviving source nodes can participate in the degraded read.

Second, source nodes are confined to nodes which contain data for reconstructing the requested data, but the starter node is not limited by this confinement. In production environments, nodes are commonly shared by a mix of application workloads. In large scale distributed storage systems, some nodes actually have much more available bandwidth than other nodes because of the load imbalance~\cite{Atikoglu12,googledata13,Huang14,Novakovic16,Lu2017Imbalance}. For example,  in Alibaba cluster, some idle servers are reserved for online service~\cite{Lu2017Imbalance}. In Haystack~\cite{Beaver2010Haystack}, different types of storage nodes co-exist, workloads of read-only nodes are much lower than that of write-enabled nodes. Therefore, we can choose a light-loaded storage node as the starter node of the degraded read. The network bandwidth for receiving data on the starter node is increased if a light-loaded starter node is chosen.

Motivated by the above analysis, we propose our approach APLS to reduce the latency of degraded read by increasing the network bandwidth for transmitting data in degraded reads.

\section{Design of APLS}
\label{sec:de}
In this section, we first present the design of APLS, and then we use an example to illustrate how APLS works, finally we theoretically analyze the benefit of APLS.

\subsection{Overview of APLS}
According to our analysis in Section~\ref{subsec:odr}, existing methods followed the way to exploit the reconstruction parallelism inside degraded reads fail to fully exploit surviving source nodes. We propose APLS (\emph{All Parallelism with a Light-loaded Starter, reads as A Plus}) to further improve the performance of degraded reads. By utilizing all surviving source nodes to do the reconstruction in parallel and assigning light-loaded starter nodes of degraded reads, APLS increases the network bandwidth for transmitting the data in degraded reads, thus reduces their latency.

According to the analysis in Section~\ref{subsec:odr}, the downstream network bandwidth of the starter node becomes the bottleneck of degraded reads. When a user request arrives, if the requested node is unavailable, a light-loaded storage node is assigned as the starter node of the degraded read to mitigate the bandwidth competition.

In RS(k,m)-coded distributed storage systems, to read an unavailable chunk, there are at least $k$ and up to $k+m-1$ source nodes which contain surviving chunks for reconstruction. To leverage all these source nodes, words of the requested chunk can be reconstructed from different groups of $k$ surviving chunks. As a result, all surviving source nodes can participate in the degraded read. Because the upstream network bandwidth of any source node may be the bottleneck, each source node sends out the same amount of data under the assumption that all source nodes have the same available network bandwidth for the degraded read.

Moreover, similar to existing methods, agent nodes are deployed on all surviving source nodes to leverage the downstream network bandwidth of source nodes. Each agent node receives the same amount of data. Because the amount of data sent out by the source nodes is more than that received by the agent nodes, the downstream network bandwidth of the agent nodes is not the bottleneck under the assumption that source nodes have the same network bandwidth for reconstruction.

Figure~\ref{fig:12} demonstrates an example of degraded reads using APLS. A light-loaded storage node is assigned as the starter node. There are five agent nodes, each of which reconstructs fifth words of chunk $C_0$. The first fifth words are reconstructed by the words from chunks $C_1$, $C_2$, $C_3$ and $C_4$, the second fifth words are reconstructed by the words from chunks $C_2$, $C_3$, $C_4$ and $C_5$, and so on. Therefore, all the five surviving source nodes participate in the degraded read, each source node sends $\frac{4\times c}{5}$ data, and each agent node receives $\frac{3\times c}{5}$ data ($c$ is the chunk size.).

\subsection{Detailed Design of APLS}
\label{subsec:ddes}
\begin{figure}[t]
	\small	
	\centering
	\includegraphics[width=.46\textwidth]{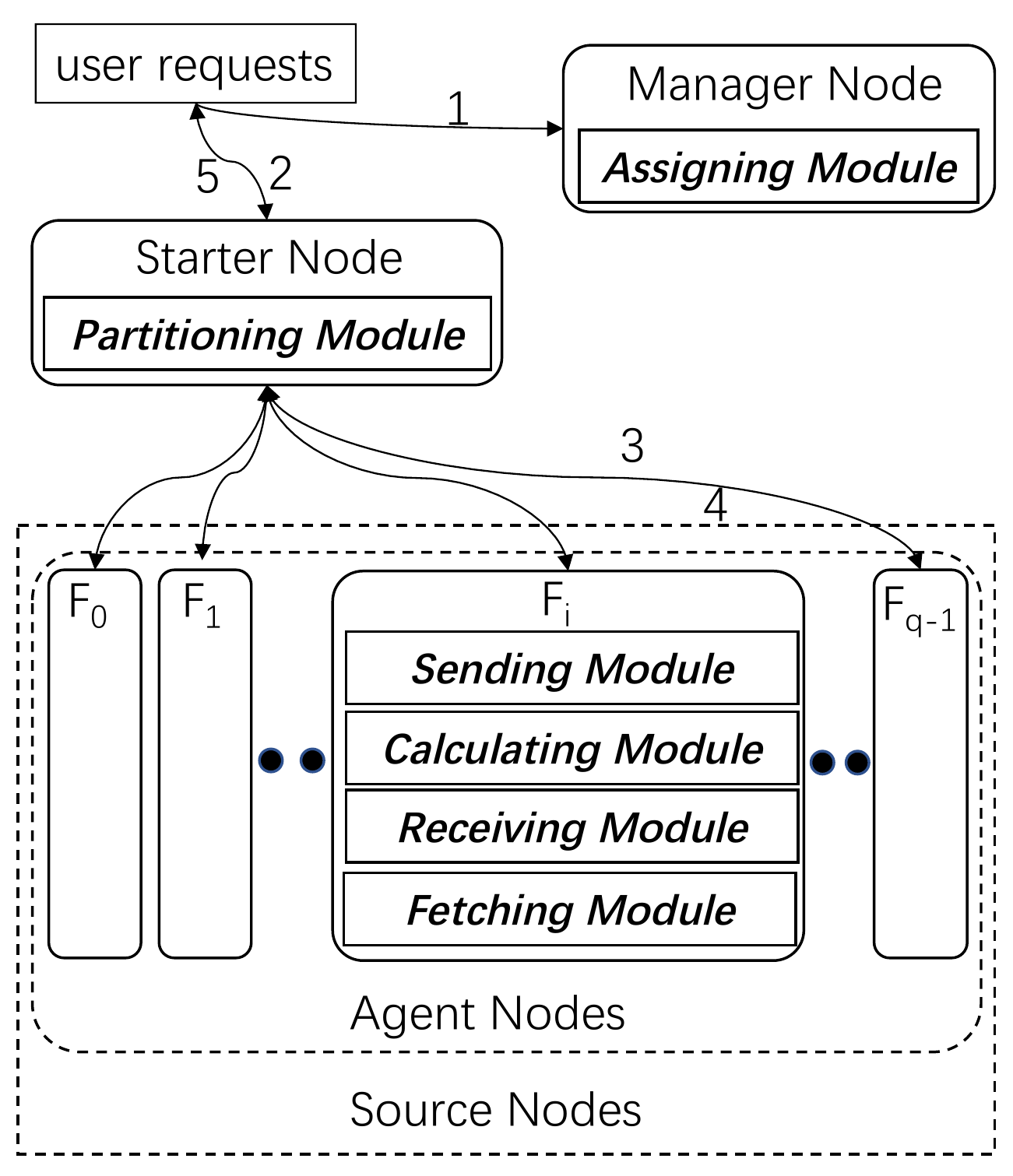}
	\caption{Architecture of APLS. The assigning module assigns starter nodes with light-loaded storage nodes based on the request history within a certain window. The partitioning module partitions the request into different agent nodes and receives requested data from the agent nodes. The fetching module directly retrieves data from the local disks, the receiving module receives data from other agent nodes over the network, the calculating module performs decoding computation on the data, and the sending module sends the data to other agent nodes or the starter node.}
	\label{fig:dor}
\end{figure}

Figure~\ref{fig:dor} demonstrates the overall architecture of APLS. When a user request arrives, a light-loaded storage node is assigned as the starter node by the manager node if the requested node is unavailable.

All surviving source nodes participate in the degraded read and are assigned as agent nodes. The starter node partitions the request into different sub-requests. The agent nodes are used to reconstruct the sub-requested data in parallel and send the requested data to the starter node.

Moreover, because the word size is too small to be efficiently transmitted over the network, a packet which contains multiple contiguous words is used. The requested chunk is split into packets, and each packet is reconstructed by one of these agent nodes. Through equally partitioning all packets on a agent node, the data for reconstruction is balanced among those agent nodes, and the requested data are transmitted by all the agent nodes. Therefore, the network bandwidth for transmitting data in the degraded read increases.

\normalsize

\subsubsection{Choosing Light-loaded Starter Node}
A statistical approach based on the request history in near past is used to choose starter nodes. Because it is hard to track the instantaneous workloads of all nodes in real time, the assigning module in the the manager node tracks a table which contains request statistics of each node measured within a certain window. The request statistics conclude the size and the corresponding nodes of all requests during the window. Then, the table is computed periodically to generate a set of nodes which have either few requests or small size of all requests. Therefore, the set contains nodes with light workloads.

When a user request arrives, if the requested node is not available, the assigning module randomly chooses a node from the set as the starter node. Then, the light-loaded starter node initiates a degraded read operation.

\subsubsection{Partitioning Requests}
In an RS(k,m)-coded distributed storage system, to read an unavailable chunk, there is $q$ (at least $k$ and at most $k+m-1$) agent nodes:  $F_0$, $F_1$, ..., and $F_{q-1}$.

\emph{Partitioning Module.} The partitioning module partitions the requested chunk into multiple sub-requests. Each agent node receives one sub-request and reconstructs part ($\frac{1}{q}$) of the requested chunk. The partitioning module sends sub-request commands to agent nodes. The sub-request command includes the start position, the read length, the code parameters, the total number of agent nodes used in this degraded read, locations of agent nodes, indexes of chunks on agent nodes, index of the unavailable chunk, a reconstruction method and the packet size.

The reconstruction method can be either the traditional method, PPR method or ECPipe method.

\begin{figure}[t]
	\centering
	\includegraphics[width=.42\textwidth ]{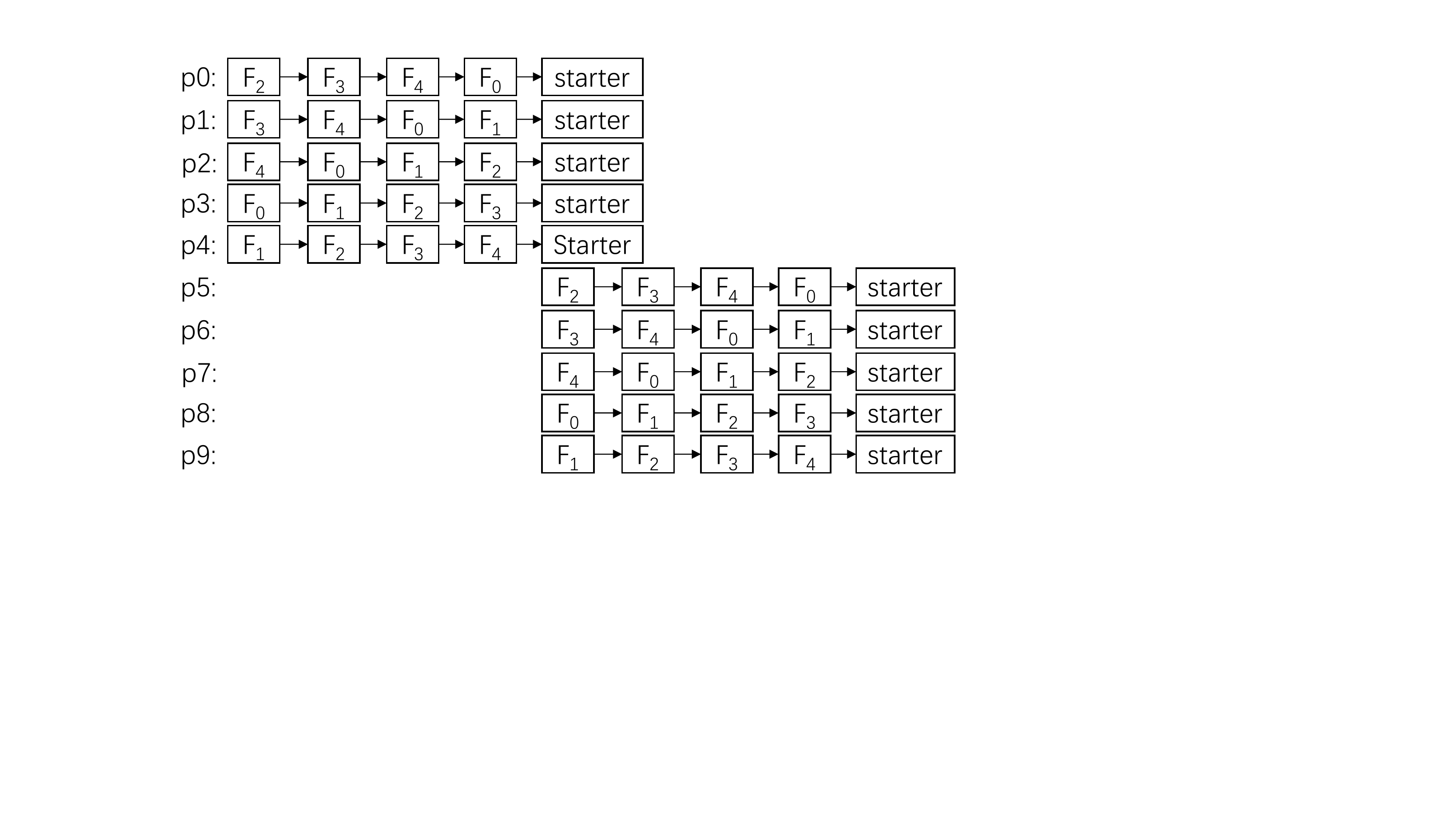}
	\caption{APLS with ECPipe method in an RS(4,2)-coded distributed storage system. For each round, 5 packets are reconstructed concurrently. Each packet is reconstructed using ECPipe method.}
	\label{fig:decpipe}
\end{figure}

\subsubsection{Reconstruction in Parallel}
\label{subsubsec:rip}
We partition the agent nodes into $q$ different \emph{reconstruction lists}, which are lists of $k$ agent nodes. Each reconstruction list $r_i$ is used to reconstruct packets for $F_i$, then $F_i$ sends the reconstructed packets to the starter node.

The composition of reconstruction lists has two rules:
(1) Each reconstruction list contains $k$ agent nodes.
(2) Each agent node belongs to $k$ different reconstruction list. Therefore, each node is response for transmitting the same amount of data during the degraded read. For example, the reconstruction list $r_i$ is composed of $k$ agent nodes: \{$F_{(i-k+1)\%q}$, $F_{(i-k+2)\%q}$, ..., $F_{i\%q}$\} ($i = 0,1,..,q-1$).

Corresponding to each reconstruction list, there is a coefficient list which includes decoding coefficients used to reconstruct packets. The coefficient list is calculated according indexes of agent nodes, the index of unavailable packet, the coding parameters (k,m) and the generator matrix. The calculation is similar to calculate the decoding matrix in Section~\ref{subsec:rsc}.

The requested chunk is split into a set of $d$ packets: $p_0$, $p_1$, ..., and $p_{d-1}$. Assumes that $p_i = \sum_{l=0}^{k-1}b_{jl}\times p_{i((j-k+l+1)\%q)}$, where $j = i\%q$ and $p_{i((j-k+l+1)\%q)}$ is the corresponding packet which is retrieved from $F_{(j-k+1+l)\%q}$, and $b_{jl}$'s are decoding coefficients. The reconstruction of the packet $p_i$ is described as follows.

\emph{Fetching Module.}
The fetching module in an agent node retrieves the data for reconstruction from local disks. To reduce the I/O competition on local disks, when an agent node receives the sub-request command from the starter node, the fetching module reads the whole chunk packet by packet and caches it in the memory.

\emph{Receiving Module.}
The receiving module in an agent node receives packets from other agent nodes. For example, by applying the traditional method, $F_j$ receives the packet $p_{i((j-k+l+1)\%q)}$ from $F_{(j-k+1+l)\%q}$.

\emph{Calculating Module.}
The calculating module calculates coefficient lists when the sub-request arrives. When a packet is retrieved by the fetching module and the corresponding packets are received over the network, the calculating module performs the decoding operation according to the decoding coefficients in coefficient lists.

For example, by applying the traditional method, $\sum_{l=0}^{k-1}b_{jl}\times p_{i((j-k+l+1)\%q)}$ are calculated on $F_j$ to get the requested packet $p_i$.

\emph{Sending Module.}
When the packet is generated by the calculating module, the sending module sends the packet to a next agent node if the packet is partially reconstructed or the starter node if the packet is fully reconstructed. For example, by applying the traditional method, $F_{(j-k+1+l)\%q}$ sends the packet $p_{i((j-k+l+1)\%q)}$ to $F_j$ and $F_j$ sends the packet $p_i$ to the starter node.

In summary, each agent node receives data of size $\frac{(k-1)\times c}{q}$, sends data of size $\frac{(k-1)\times c}{q}$ to other agent nodes, and sends data of size $\frac{c}{q}$ to the starter node ($c$ is the chunk size.). Therefore, the network transmission is balanced in the degraded read.

\subsubsection{Reconstruction using Pipelining}
\label{subsubsec:iwe}
Besides reconstructing packets of unavailable chunks in parallel in Section~\ref{subsubsec:rip}, those packets can be also reconstructed using pipelining. ECPipe~\cite{ATC17-lirepair} uses repair pipelining to reconstruct an unavailable chunk as shown in Figure~\ref{fig:p3}. Figure~\ref{fig:decpipe} shows the data flow of APLS with ECPipe method.


As shown in Figure~\ref{fig:decpipe}, in an RS(4,2)-coded distributed storage system, there are 5 surviving agent nodes. They are split into 5 reconstruction lists: \{$F_2$, $F_3$, $F_4$, $F_0$\}, \{$F_3$, $F_4$, $F_0$, $F_1$\}, \{$F_4$, $F_0$, $F_1$, $F_2$\}, \{$F_0$, $F_1$, $F_2$, $F_3$\} and \{$F_1$, $F_2$, $F_3$, $F_4$\}. For example, packet $p_8$ is reconstructed by $r_3$. Assume that $p_8 = \sum_{l=0}^{3}b_{3l}\times p_{8l}$, where $p_{8l}$ is retrieved from $F_l$. At the first step, $F_0$ sends $b_{30}\times p_{80}$ to $F_1$; at the second step, $F_1$ sends $\sum_{l=0}^{1}b_{3l}\times p_{8l}$ to $F_2$; at the third step, $F_2$ sends $\sum_{l=0}^{2}b_{3l}\times p_{8l}$ to $F_3$; finally, $F_3$ sends $\sum_{l=0}^{3}b_{3l}\times p_{8l}$ to the starter node. Moreover, packets $p_0$, $p_1$, $p_2$, $p_3$ and $p_4$ are reconstructed concurrently.

\subsection{Theoretical Analysis}
\label{subsec:mod}
In this section, we illustrate the benefit of APLS by quantitatively analyzing the latency of degraded reads in an RS(k,m)-coded distributed storage system.

\emph{Model Assumption.}
We assume that source nodes have the same network bandwidth for reconstruction. Let $\theta_s$ denote the ratio of the network bandwidth available for the degraded read to the total network bandwidth on source nodes. We assume all source nodes have the same available network bandwidth for the degraded read.

We assume that $q$ ($q \ge k$) source nodes participate in a degraded read. Agent nodes are used to reduce the downstream network bandwidth of the starter node. In order to reduce the network transmission, agents nodes are co-located with source nodes.

The latency of the degraded read is most affected by the network bandwidth~\cite{Eurosys2016PPR, ATC17-lirepair}, the decoding computation and disk I/O are neglected. Therefore, we discuss the theoretical latency of the degraded read from the view of data transmission.

\textbf{\emph{Case 1: The starter node is chosen from source nodes.}}

One of the $q$ source nodes is assigned as the starter node. Because the requested chunk is reconstructed on the starter node, the starter node needs to receive $c$ data.

In order to utilize the downstream bandwidth of source nodes, agent nodes are deployed with the rest $q-1$ nodes. The agent nodes receive $(k-2)\times c$ and send $(k-1)\times c$ data: ($(k-2)\times c$ data to other agent nodes and $c$ data to the starter node). Therefore, the downstream bandwidth of the agent nodes is not the bottleneck. Because each agent node sends $\frac{(k-1)\times c}{q}$ ($\le c$) data, the upstream bandwidth of the agent nodes is not the bottleneck.

When the starter node is chosen from source nodes, the downstream bandwidth of the starter node is the bottleneck. Therefore, the latency of the degraded read is calculated by Equation~\eqref{eqa:tdr1}.

\begin{equation}
\label{eqa:tdr1}
T_{dr}(c) = \frac{c}{\theta_s\times B}
\end{equation}

According to Equation~\eqref{eqa:tdr1}, simply utilizing more source nodes ($> k$) cannot further reduce the latency of the degraded read because the downstream network bandwidth of the starter node is the bottleneck.

\textbf{\emph{Case 2: The starter node is not chosen from source nodes.}}
By using a light-loaded starter node, we assume that the starter node can utilize all available network bandwidth in the degraded read.

The $q$ source nodes are agent nodes. Because the reconstructed chunk is sent from the agent nodes to the starter node, the starter node needs to receive $c$ data.

The agent nodes receive $(k-1)\times c$ and send $k\times c$ data ($(k-1)\times c$ data to other agent nodes and $c$ data to the starter node). Therefore, the downstream network bandwidth of the agent nodes is not the bottleneck. Because each agent node sends $\frac{k\times c}{q}$ ($\le c$) data and the starter node receives $c$ data, the downstream network bandwidth of the starter node is not the bottleneck when we choose a light-loaded starter node. Therefore, the latency of the degraded read is calculated by Equation~\eqref{eqa:tdr}.

\begin{equation}
\label{eqa:tdr}
T_{dr}(c)=\frac{k \times c}{q\times\theta_s\times B}
\end{equation}

According to Equation~\eqref{eqa:tdr}, $T_{dr}(c)$ decreases when $q$ goes up. Although RS codes can tolerate multiple simultaneous failures, single failure is the most common case~\cite{HOTSTORAGE13Rashmi-warehouse, fds-osdi12, USENIX12HUANG-LRC}. Thus, the number of source nodes is typically $k+m-1$. According to Equation~\eqref{eqa:tdr1} and Equation~\eqref{eqa:tdr}, because $q$ can be larger than $k$, the latency of the degraded read optimized by APLS is lower than that of the degraded read optimized by the state-of-the-art solutions.
  
For normal read, we assume the requested node have the same available network bandwidth with source nodes, therefore, the latency of the degraded read is $\frac{c}{\theta_s\times B}$. According to Equation~\eqref{eqa:tdr}, the latency of degraded reads can be lower than that of normal reads in theory if $q$ is larger than $k$.

\section{Evaluation}
\label{sec:pe}
We implement a prototype to verify the correctness and effectiveness of the design of APLS in real world environments. Our prototype named \emph{APLS} is implemented upon ECPipe~\cite{lecpipe}, an open source storage system that supports RS coding. Our methods are general enough to be applicable to other RS coded storage systems. We choose ECPipe for two reasons: it has simple architecture and is well optimized for degraded reads.

\begin{figure*}[t]
	\centering
	\includegraphics[width=.96\textwidth ]{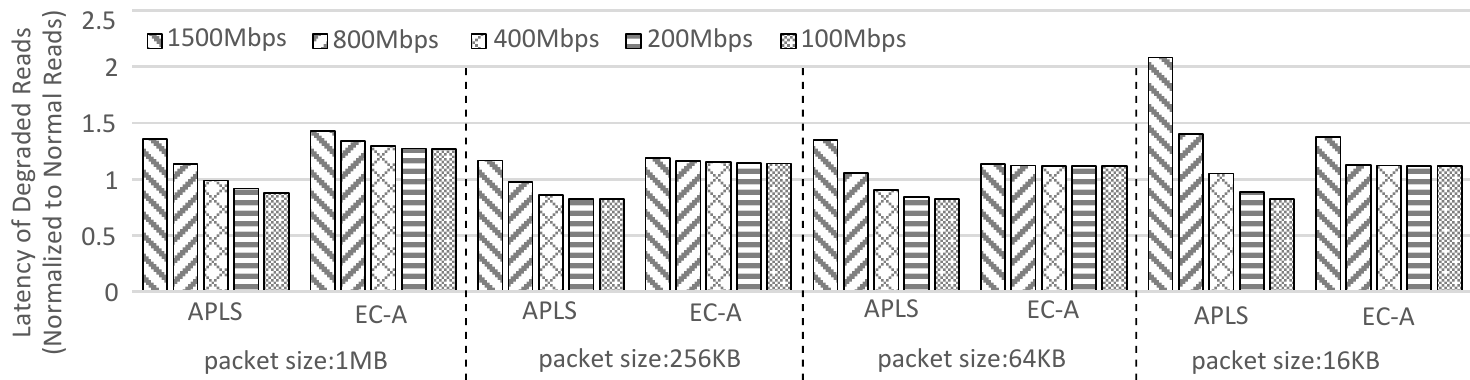}
	\caption{Impact of different packet sizes on the latency of degraded reads. The chunk size is 64 MB and the packet size changes from 16 KB to 1 MB.  The RS(10,4) coding scheme is used. The network bandwidths of helpers are limited from 100 Mbps to 1500Mbps, which indicates the background workloads range from heavy to light.}
	\label{fig:res1}
\end{figure*}

ECPipe consists of three major components: a centralized coordinator, helpers, and requestors. We implemented APLS on ECPipe by modifying these three components according to the architecture in Figure~\ref{fig:dor} and the data flow in Figure~\ref{fig:decpipe}. We add about 400 lines of codes.

The requestor refers to an entity that interfaces with the user requests, it sends a read request with the identifier of the unavailable chunk to the coordinator and receives data from helpers. APLS implements the partitioning module on the requestor.

The centralized coordinator manages locations of RS-coded chunks and the reconstruction operation between a requestor and multiple helpers. In ECPipe, the coordinator uses the identifier of the unavailable chunk to find locations of only $k$ surviving chunks in the same stripe. First, the coordinator in APLS finds locations of all surviving chunks in the same stripe. Second, APLS adds a new command which is sent to helpers and sends the command to all surviving helpers. The command includes the start position, the read length, the code parameters, the total number of surviving helpers, locations of surviving helpers, indexes of chunks on helpers, index of the unavailable chunk, and the packet size.

The helper runs on each server where the data are hosted, and manages disk I/O. First, APLS adds the calculating module to each helper to calculate decoding coefficients and perform calculations. Second, APLS modifies the sending and the receiving procedure to implement the order of sending and receiving packets as the order in Figure~\ref{fig:decpipe}.

\subsection{Experimental Setups}
Our testbed is an RS-coded storage cluster that consists of sixteen ECS servers running on the Alibaba Cloud~\cite{aliyun}. Each server contains an Intel Xeon E5-2682 v4 @ 2.5 GHZ CPU, 8 GB DDR3 memory, 1.5 Gbps network and 40GB SSD. The operation system running on all these servers is Ubuntu 14.04. The coordinator runs on one server, the requestor runs on another server, and 14 helpers run on the remaining servers.

Our experiments mainly evaluate the latency of degraded reads. For a degraded read, the latency is from the time when a requestor send a read request to the time when it finish the reconstruction of the unavailable chunk. For each experiment, the mean values is computed from 10 runs. Our goal is to bridge the gap between normal reads and degraded reads. Thus, the results are normalized to normal reads (which read the same amount of data directly from the target server).

Comparisons between APLS and ECPipe are made under the scenarios described as follows.

(1) Because nodes are affected by background workloads, we use the Linux command tc~\cite{linuxtc} to limit the network bandwidth for degraded reads. Because Alibaba Cloud provides at least 1.5 Gbps network bandwidth for each server, we first limit the total network bandwidth of each server to 1500 Mbps. In below experiments, to illustrate the effect of assigning light-loaded starter nodes, the network bandwidth of the server which hosts the requestor keeps the 1500 Mbps limitation. \textbf{The network bandwidth of servers which host the helpers are limited from 100 Mbps ($\delta_s$ = $0.067$) to 1500Mbps ($\delta_s$ = $1$), which indicates the background workloads range from heavy to light}.

(2) Because the requested data are transmitted packet by packet, the affect of the packet size is studied. The packet size changes from 16KB to 1MB. The chunk size is configured as 64 MB, and the RS(10,4) coding scheme~\cite{HOTSTORAGE13Rashmi-warehouse} is used.

(3) To study the effect of APLS on small sized chunks, the chunk size is configured as 256 KB and 4 MB~\cite{atc17-giza}. The RS(10,4) coding scheme is used.

(4) From our analysis in Section~\ref{subsec:mod}, the number of source nodes has a significant effect on the performance of degraded reads. To study the affect, the RS(6,6) coding scheme is used. Then, the number $q$ of source nodes changes from $6$ to $11$. \textbf{It is worth noting that the effect of different $q$ can also be seen as the effect of different RS coding schemes}.

Besides, ECPipe have two versions, EC-A and EC-B, they both use $k$ helpers to serve degraded reads, the difference is that EC-A uses one helper to send the requested data to the requestor and EC-B uses $k-1$ helpers to send the requested data to the requestor. EC-A is measured in all experiments and EC-B is measured only in the experiments in Section~\ref{subsubsec:ps}.

\begin{figure*}[t]
	\centering
	\includegraphics[width=.96\textwidth ]{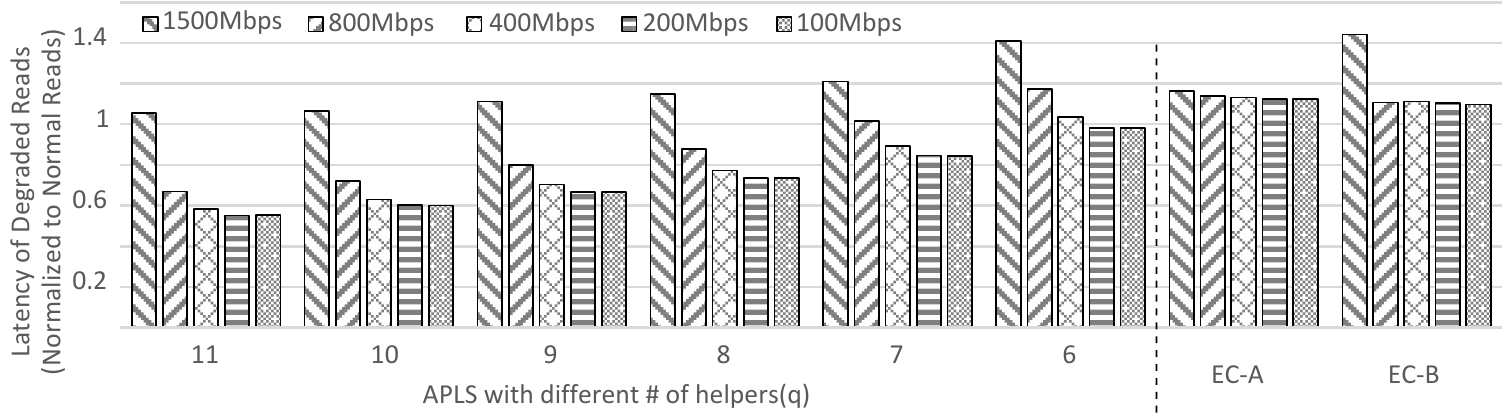}
	\caption{Impact of different numbers of helpers on the latency of degraded reads. The chunk size is 64 MB. The RS(6,6) coding scheme is used. The network bandwidths of helpers are limited from 100 Mbps to 1500 Mbps, which indicates the background workloads range from heavy to light. EC-A uses one helper to send the requested data to the requestor and EC-B uses $k-1$ helpers to send the requested data to the requestor.}
	\label{fig:res3}
\end{figure*}

\begin{figure}[t]
	\centering
	\includegraphics[width=.46\textwidth ]{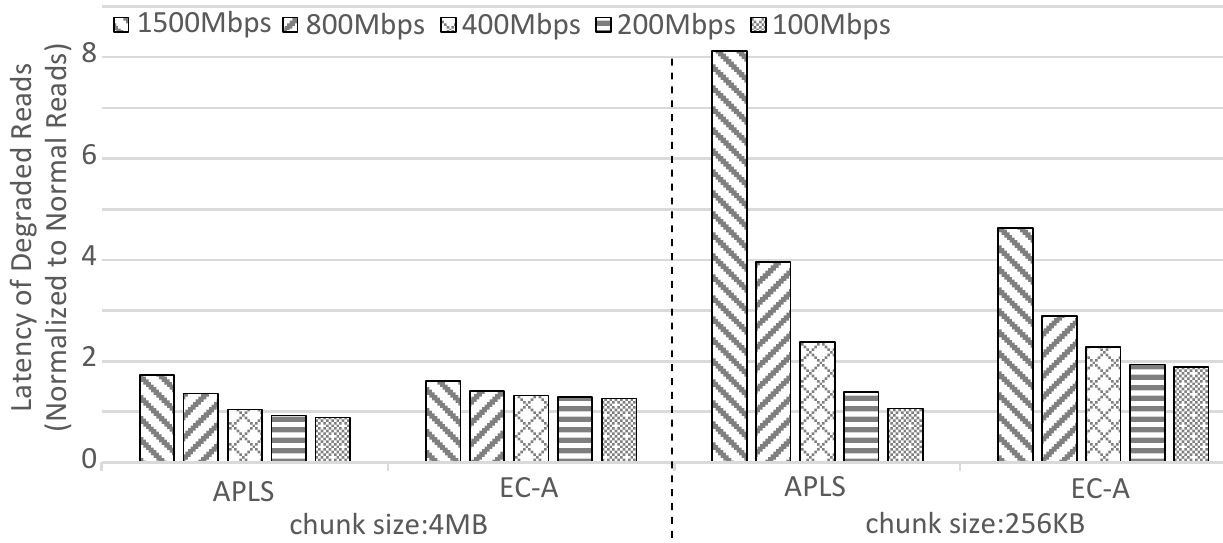}
	\caption{Impact of different chunk sizes on the latency of degraded reads. The chunk sizes are 4 MB and 256 KB. The RS(10,4) coding scheme is used. The network bandwidths of helpers are limited from 100 Mbps to 1500Mbps, which indicates the background workloads range from heavy to light.}
	\label{fig:res2}
\end{figure}

\subsection{Experimental Results}
\label{subsec:er}

\subsubsection{Impact of Packet Size}
\label{subsubsec:ps}
We first evaluate the affect of the packet size. The latency of normal reads is limited the network bandwidth of both the helper and the requestor. \textbf{For example, when the network bandwidth of the helpers is limited 100 Mbps, the time of directly reading a 64 MB chunk is 5.12 seconds.} The latency of degraded reads is normalized to that of normal reads.

Five observations are derived from Figure~\ref{fig:res1}. First, in most cases, APLS performs better than ECPipe. The improvement of APLS over ECPipe becomes more significant when the helpers have heavier background network workloads. For example, using the packet size of 64 KB, compared to ECPipe, APLS reduces the latency by 6\% to 25\% when the network bandwidth limitations of the helpers are changed from 800 Mbps to 100 Mbps.

Second, in most cases, APLS achieves better performance over normal reads. The improvement of APLS over normal reads becomes more significant when the helpers have heavier background network workloads. This is consistent with our analysis in Section~\ref{subsec:mod}. For example, using the packet size of 256 KB, compared to normal reads, APLS reduces the latency by 3\% to 17\% when the network bandwidth limitations of helpers are changed from 800 Mbps to 100 Mbps.

Third, APLS has the lowest latency when the packet size is 256 KB, while ECPipe has the lowest latency when the packet size is 64 KB. When the helpers have no background workload, which happens infrequently, ECPipe is a little better than APLS. For example, when the network bandwidth limitations of the helpers are 1500 Mbps, compared to ECPipe with the packet size of 64 KB, APLS with the packet size of 256 KB increases the latency by only 3\%. The reason is that, under such a scenario, the latency is limited by the network bandwidth of the requestor. Because the requestor receives the requested data from ten helpers, the competition of the downstream network traffic worsens the performance of APLS.

Fourth, when the packet size becomes smaller than 64 KB, both APLS and ECPipe have higher latency. The reason is that too many packets are transmitted over the network, which significantly affect the performance of the network transmission.

Fifth, in some cases, especially under heavy workloads, the latency of degraded reads is lower than that of normal reads. It is consistent with our analysis in Section~\ref{subsec:mod}.

\subsubsection{Impact of Chunk Size}
\label{subsubsec:ssc}
To study the effect of APLS on small sized chunk, the chunk size is configured as 4 MB and 256 KB. The latency of degraded reads is normalized to that of normal reads.

We have two observations. First, APLS is still effective when the chunk size is small and the helpers have heavy background network workloads. For example, when the chunk size is 4 MB and the network bandwidth limitations of helpers are 100 Mbps, APLS has a 11\% lower latency than normal reads. When the helpers have light background network workloads, APLS is not as good as normal reads. The reason is that when the chunk size is small, the network hop latencies and the synchronization latencies of the pipelining have more influence on the latency of degraded reads.

Second, APLS always outperforms ECPipe even the chunk size is reduced to 256 KB. For example, compared to ECPipe, APLS reduces the latency by 28\% when the network bandwidth limitations of helpers are 200 Mbps. Only when the helpers have light background workload, ECPipe is better than APLS, the reason is explained in Section~\ref{subsubsec:ps}.

\subsubsection{Impact of The Number of Source Nodes}
\label{subsubsec:nq}
To study the effect of the number of source nodes ($q$), we use the RS(6,6) coding scheme. In APLS, $q$ is changed from 6 to 11. The latency of degraded reads is normalized to that of normal reads.

We have derived three observations from Figure~\ref{fig:res3}. First, the improvement of APLS increases when $q$ increases. Especially, APLS significantly improves the latency when the helpers have medium or heavy background network workloads. For example, compared to normal reads, when the network bandwidth limitations of helpers are 100 Mbps, the percentage reduction of the latency is from 16\% (1/7) to 45\% (5/11) when $q$ increase from 7 to 11. This is consistent with our analysis in Section~\ref{subsec:mod}. The reason is that, the bottleneck is the upstream bandwidth of the helpers, thus the latency of degraded reads is $\frac{6}{q}$ of that of normal reads.

Second, APLS achieves better performance than both EC-A and EC-B in most cases. This is also consistent with our analysis in Section~\ref{subsec:mod}.

Third, EC-B performs a little better than EC-A when helpers affected by background workloads. The reason is that, under such a scenario, the latency is limited by the upstream network bandwidth of the helpers. Because EC-B uses 5 helpers to send the requested data to the requestor, thus improves the usage of the downstream network bandwidth of the requestor. However, when helpers are not affected by background workloads, EC-A is better than EC-B because EC-B uses five helpers to send the requested data to the requestor, the competition of the downstream network traffic of the requestor worsens the performance of EC-B.

\section{Related Work}
\label{sec:rw}
In this section, we summarize existing solutions which improve the performance of degraded reads.

Many solutions have been proposed to reduce the latency of degraded reads by reducing data transmission for reconstruction~\cite{tott2010network, Greenan2010Flat, SIGMETRICS10Optimal, USENIX12HUANG-LRC, Khan2012EC, fast2012NCCloud, TOTT2013Zigzag, VLDB2013XORing, HOTSTORAGE13Rashmi-warehouse, TOC2014Single, Rashmi2014hitchhiker, Xia2015tale, FAST16butterfly}. Both new MDS (Maximum Distance Separable) and non-MDS erasure codes~\cite{Greenan2010Flat, fds-osdi12, USENIX12HUANG-LRC, Khan2012EC, fast2012NCCloud, TOTT2013Zigzag, VLDB2013XORing, HOTSTORAGE13Rashmi-warehouse, eurosys2014archiving, Xia2015tale, Rashmi2015Having, FAST16butterfly} are designed to reduce the data transmission for reconstruction. Regenerating Codes~\cite{tott2010network,SIGOPS2013RCS, eurosys2014archiving} are a family of MDS codes. The data transmission of reconstruction in degraded reads of the regenerating codes is much lower than that of traditional RS codes. However, the regenerating codes are not systematic codes, thus suffer from high read cost. To maintain low data transmission of reconstruction and read cost, systematic MDS codes, such as Zigzag and Butterfly codes~\cite{TOTT2013Zigzag, FAST16butterfly}, are proposed. The application of the regenerating codes has many limitation, such as the storage overhead, and the computation cost, therefore, the regenerating codes are seldom used in production environments. Compared to the MDS codes, non-MDS codes, as LRC~\cite{USENIX12HUANG-LRC, VLDB2013XORing, Xia2015tale}, dramatically reduce the data transmission of reconstruction. However, the cost of non-MDS codes cannot be ignored, particular when the scale of the data center is very large, i.e., even 1\% extra storage overhead usually means millions of dollars~\cite{dutta2013cost, AmazonS3price}.

Besides exploiting the reconstruction parallelism inside degraded reads~\cite{Eurosys2016PPR, ATC17-lirepair}, the latency of degraded reads is reduced by exploiting the parallelism between degraded reads~\cite{TPDS2014On, toc2015zhuboost,ec15}. FastDR~\cite{toc2015zhuboost} exploits I/O parallelism between degraded reads to improve the performance of degraded reads in heterogeneous distributed storage systems. EDP~\cite{Shen2016Encoding} designs an order to generate parity elements by using data elements and refines the data layout to reduce the number of I/Os in degraded reads during a single failure, thus reducing the read latency.

There have been extensive studies on reducing the tail latency in erasure-coded storage systems~\cite{Gardner2015Reducing, Hu2017Latency}. Hu et al.~\cite{Hu2017Latency} propose proactive degraded reads and load balancing mechanisms to reduce the median and tail latencies of reads. There also have been other studies to reduce the tail latency of degraded reads by sending abundant chunks~\cite{Gardner2015Reducing, AggarwalAFL17}. To get an unavailable chunk in an RS(k,m)-coded storage system, the client receives more than $k$ chunk. The first $k$ chunks are used to reconstruct the unavailable chunk. This scheme is mainly used by non-systematic codes. For example, Aggarwal et al.~\cite{AggarwalAFL17} proposes a latency tail probability model to minimize the tail latency of all files by optimizing the chunk placement and the scheduling policy.

\section{Conclusions}
\label{sec:con}
In this paper, we propose a parallel reconstruction method, called APLS to improve the performance of degraded reads. APLS utilizes all surviving source nodes to reconstruct unavailable chunks in fine-granularity and assigns light-loaded starter nodes to receive the requested data. Therefore, APLS increases the network bandwidth for transmitting the data of reconstruction, thus reducing the latency of degraded reads. Both modeling and prototyping are used to verify the effectiveness and efficiency of APLS. Experimental results collected from prototyping which is built on a storage cluster of sixteen servers demonstrate that, compared to the state-of-the-art solution - ECPipe, APLS further reduces the latency of degraded reads by up to 28\% when nodes have medium or heavy workloads.

{\footnotesize \bibliographystyle{IEEEtran}
	\bibliography{IEEEabrv,rafibib}}

\end{document}